\documentclass[aps,prd,longbibliography,twocolumn,nofootinbib,floatfix]{revtex4-1}
\usepackage{setspace}
\usepackage{amsthm}
\theoremstyle{definition}

\usepackage{graphicx}
\usepackage{dcolumn}
\usepackage{multirow}
\usepackage{subfigure}
\usepackage{times,mathptm}
\usepackage{float}
\usepackage{color}
\usepackage{bbold}
\usepackage{amsmath,amsfonts,amssymb}
\usepackage{mathptmx}
\usepackage{mathrsfs}
\usepackage{bbm}
\usepackage{bm}
\usepackage{xfrac}

\newcommand{\beq}{\begin{equation}}
\newcommand{\eeq}{\end{equation}} 
\newcommand{\bea}{\begin{eqnarray}}
\newcommand{\eea}{\end{eqnarray}} 
\newcommand{\op}{\overline{\phi}}
\newcommand{\Om}{\Omega}
\newcommand{\tvp}{\widetilde{\varphi}}

\newcommand{\U}{{\cal U}}

\newcommand{\Ac}{{\cal A}}

\renewcommand{\d}{\delta}

\renewcommand{\l}{\lambda}

\newcommand{\Q}{\mathbf{Q}}

\newcommand{\tphi}{\tilde{\phi}}

\renewcommand{\b}{\beta}
\renewcommand{\a}{\alpha}

\newcommand{\tr}{\text{Tr}}

\newcommand{\bx}{\mathbf{x}}
\newcommand{\by}{\mathbf{y}}

\newcommand{\vphi}{\varphi}
\newcommand{\vx}{{\vec{x}}}
\newcommand{\vy}{{\vec{y}}}
\newcommand{\vz}{\vec{z}}

\newcommand{\bU}{{\bf U}}

\newcommand{\m}{\mu}
\newcommand{\pbar}{\overline{\psi}}

\newcommand{\g}{\gamma}
\renewcommand{\r}{\rho}

\newcommand{\s}{\sigma}
\renewcommand{\k}{\kappa}

\newcommand{\tU}{\widetilde{U}}

\renewcommand{\th}{\theta}

\newcommand{\vph}{\varphi}
\newcommand{\oh}{\frac{1}{2}}

\newcommand{\dg}{\dagger}
\newcommand{\non}{\nonumber}

\newcommand{\rf}[1]{(\ref{#1})}
\newcommand{\ra}{\rightarrow}
\newcommand{\pa}{\partial}
\renewcommand{\vec}[1]{\bm #1}

\usepackage{ulem}

\begin{document}

\title{Varieties of electrically charged physical states in SU(2)$\times$U(1) lattice gauge Higgs theory}
\bigskip
\bigskip

\author{Jeff Greensite}
\affiliation{Physics and Astronomy Department \\ San Francisco State
University  \\ San Francisco, CA~94132, USA}
\bigskip
\date{\today}
\vspace{60pt}
\begin{abstract}

\singlespacing

We consider a quenched SU(2)$\times$U(1) gauge Higgs theory on the lattice, coupled to a static vector-like fermion which, in this case, is in the same gauge group representation as the Higgs field.   Physical (i.e.\ locally gauge invariant) electrically charged and electrically neutral states of matter particles in the electroweak theory were described decades ago, but those constructions do not exhaust all the possibilities, and new types of electrically charged/neutral states, orthogonal to former constructions, are described here.  The difference has to do with how the static source, which by itself does not create a physical state, is dressed by dynamical fields.   We find that, unsurprisingly, the neutral static fermion is much lighter than any of the charged fermion states.  But a lattice study of the propagation of the charged fermion states indicates the existence of (at least) two particle 
states with different masses in charged particle spectrum.

\end{abstract}

%
%
%
\maketitle
 
\singlespacing

\section{\label{Intro} Introduction}

    Gauss's Law tells us that charged particles are the source of electric fields, and in quantum gauge field theory the Gauss law becomes a requirement of local gauge invariance for all physical particle states.  Such states can be regarded as consisting of one of more charged particles which are dressed in some way by gauge and (depending on the theory) other fields as well.  For a particle-antiparticle pair in the confining phase, the dressing takes the form of a color electric flux tube, at least up to the string breaking distance.  There are physical flux tube states beyond that breaking distance which are metastable. In a Coulomb phase, the dressing for static particles has the form of a Coulomb field. But the nature of electrically charged physical states in the Higgs phase of an SU(2)$\times$U(1) theory is a little subtle.  Of course states of this kind were constructed decades ago by  `t Hooft \cite{tHooft:1979yoe}, Banks and Rabinovici \cite{Banks:1979fi}, and Fr{\"o}hlich, Morchio, and Strocchi \cite{Frohlich:1981yi}. I will argue that there
exist other, qualitatively different states in this theory which were not covered in the old work, and which ought be taken into account in a spectrum calculation. 

     In quantum mechanics, a composite bound state generally has a spectrum of energy eigenstates.  Atoms, nuclei, and hadrons come to mind.  So does the gluelump in QCD, where a static particle in the adjoint representation of SU(3) color is bound to, and neutralized by, a surrounding gluon field, and many studies have shown that there exists a discrete spectrum of gluelump physical states, see e.g. \cite{Herr:2023xwg,Guo:2007sm,Bornyakov:2001qw}.  With the example of the gluelump in mind, this brings up an interesting question:  What about static vector-like fermions in an SU(2)$\times$U(1) gauge Higgs theory?  These are also systems in which electroweak gauge and Higgs fields surround a charged source, so they are composite particles in the same sense that gluelumps are composite particles.  Then do these particles also have a spectrum of excitations? The energy of interest is energy above the large bare mass of the static particle.  At this stage we make no claims about what happens in the realistic electroweak theory, which would require a formulation of chiral fermions on the lattice, and although there exist some interesting recent proposals \cite{Kaplan:2024ezz,Thorngren:2026ydw,none},  their practicality in numerical simulations is not yet clear.  So for now we restrict the calculation to vector like particles, taken to be extremely massive for simplicity.  Lattice couplings in the SU(2)$\times$U(1) theory are also chosen for convenience.
   
     Before calculating excitations of a static fermion of this kind, it is necessary to first describe the electrically
neutral (``neutrino'') and electrically charged (``electron'') physical states, which must each be locally gauge invariant.  This was done in part, and long ago, by the authors cited above \cite{tHooft:1979yoe,Banks:1979fi,Frohlich:1981yi}, but I will argue that none of these provides a complete listing of the possibilities, and that in fact there are two distinct types of both charged and neutral physical states which were not fully described in the old work, and which ought be taken into account.  The difference I am referring to concerns the distinct ways in which a matter source can be dressed by the gauge and Higgs fields in order to form physical states with the same electric charge, but differing transformation properties under a global subgroup of the SU(2)$\times$U(1) gauge group.

     Section \ref{charge} introduces the lattice action and the ``pseudomatter'' fields required for the construction of charged physical states. Section \ref{states} describes two general classes of physical states in 
SU(2)$\times$U(1) containing electrically charged and neutral fermions.  Numerical spectrum results, derived from separated fermion-antifermion  pair propagation, are presented in section \ref{results}, with conclusions in section \ref{conclude}.

\section{\label{charge}Charge in SU(2)$\times$U(1) gauge Higgs theories}

   A general question in gauge Higgs theories is whether the Higgs region of the phase diagram is
a genuine phase, sharply distinguished by some order parameter and by the spontaneous breaking of
some global symmetry from other regions of the phase diagram.  Our view was put forward in
\cite{Greensite:2020nhg,Greensite:2021fyi}, and is based on the equivalence of 
expectation values of operators at fixed time in a gauge Higgs theory with expectation values of those operators in a certain type of spin glass.  The Higgs phase is the spin glass phase, the order parameter is non-local, and the symmetry which is broken is the global center subgroup of the gauge group.  This will be reviewed in an appendix.

    However, in this article we are concerned only with the Higgs phase of the SU(2)$\times$U(1) theory, in which there still exist physical states which are either neutral or charged with respect to the electromagnetic subgroup of the gauge group U$_\text{EM}$(1) .   The strategy to construct fermionic states, following the work of \cite{tHooft:1979yoe,Banks:1979fi,Frohlich:1981yi}, is to construct locally gauge covariant operators from the fermion and scalar fields which are either electrically neutral or electrically charged under U$_\text{EM}$(1) . If the operators are covariant rather than invariant, then they must be multiplied by a covariant (and non-local) functional of the gauge fields, a ``pseudomatter" field, in order to achieve the local gauge invariance required of physical states.  

A standard example of a pseudomatter field in gauge theory is found in the physical state of a charged static fermion coupled to the quantized Maxwell field in an infinite volume \cite{Dirac:1955uv},\footnote{Infinite volume is required because it is impossible to place a single electric charge in a finite periodic volume.  To overcome this problem we will shortly consider well separated particle-antiparticle pairs.}
\beq
  \Psi_{\text{chrg}} (x;A)=  \pbar(x) \rho(x;A) \Psi_0[A] \ ,
 \label{QED}
 \eeq
 where $\Psi_0$ is the electromagnetic ground state, $\pbar$ creates a static fermion, and
 \bea
       \rho(x;A) &=&  \exp\left[-i {e\over 4\pi} \int d^3z ~ A_i(\vz) {\pa \over \pa z_i}  {1\over |\vx-\vz|}  \right] 
 \label{rho}
 \eea
is an example of a pseudomatter field.  By definition, a pseudomatter field transforms covariantly like a matter field under the gauge group, except for transformations in the global center subgroup, under which it is invariant.  In this case, the state \rf{QED} is invariant under local gauge transformations, but if we consider a global U(1) transformation $g = e^{i\th}$, then $\psi(x) \ra e^{i\th}\psi$, but $\rho(x;A)$ is invariant.  This means that the physical state $\Psi_{\text{chrg}}  \ra e^{-i\th} \Psi_{\text{chrg}}$ is covariant rather than invariant, under the global center subgroup.  In contrast, given a scalar field $\phi(x)$
of unit charge, then the electrically neutral state
\beq
  \Psi_{\text{neutral}}(x;A) =  \pbar(x) \phi(x) \Psi_0[A]  
\eeq
is invariant under all gauge transformations, including the global center transformations.  The charged and neutral states are therefore orthogonal, if the vacuum is invariant under the global symmetry.  If the vacuum
is not invariant under the global U(1) gauge group, as in the Higgs phase of the abelian Higgs model,
then this orthogonality breaks down.

For SU(2)$\times$U(1) in the Higgs phase we cannot use transformation properties in the global center subgroup to guarantee the orthogonality of electrically charged and neutral states, since that symmetry is spontaneously broken.  Such states must be orthogonal for some other reason, which we will discuss below.

To fix notation, we work in an SU(2)$\times$U(1) lattice formulation (cf.\
\cite{Shrock:1985ur,Zubkov:2008gi} for early studies) which we express as
\bea
  S  &=& - \sum_{plaq} \bigg\{ \b_1 \oh \tr[UUU^\dg U^\dg] + \b_2 \text{Re}[VVV^\dg V^\dg] \bigg\} \non \\
      & &  - 2 \sum_{x,\m} \text{Re}[\phi^\dg(x) U_\m(x) V_\m(x) \phi(x+\hat{\m})]  \non \\
      & & +\sum_x [(-\g+8)\phi^\dg(x)\phi(x) + \lambda (\phi^\dg(x)\phi(x))^2] \ ,
 \eea
 where $U,V$ are SU(2) and U(1) valued link variables, respectively,
and we will also include a static vector-like fermion in the same gauge group representation as the Higgs field, and coupled to the gauge fields in the same way.  This non-dynamical field will be treated in a hopping parameter expansion
with the hopping parameter $\k$ taken to be small enough, in lattice units, to justify treating the fermion as static, propagating only in the time direction, and fermion loops are dropped.   
 Links $U$ transform via the SU(2) gauge group, links
$V$ via U(1), and we denote the product $\tU_\m(x)=U_\m(x)V_\m(x)$.  

\subsection{Pseudomatter fields}

Pseudomatter fields have been defined previously as functionals of the gauge fields, in this case the
$U,V$ fields, transforming covariantly like a matter field in the defining representation of gauge group, 
but remaining invariant under transformations in the global center subgroup of the gauge group, since these do not transform the gauge fields.  An example of such a pseudomatter field was shown in \rf{rho} above for  the abelian theory.  Transformations to a physical gauge, defined by some condition to be satisfied by the gauge fields, can also be associated with pseudomatter fields, and in fact  $\rho^\dg(x,A)$ in \rf{rho}  is precisely the transformation of a single charged matter field to Coulomb gauge.  
It goes the other way also.  Pseudomatter fields can be used to define gauges,
cf.\ Vink and Wiese \cite{Vink:1992ys} for a gauge which avoids the Gribov problem.   

For our construction of charged states we require pseudomatter fields $\r_n(x;V)$ which are functionals of the U(1) field $V$ alone, transforming covariantly via the U(1) subgroup, and also pseudomatter fields $\xi_n(x;\tU)$ which are functionals of the $\tU=UV$ field, transforming covariantly under the full SU(2)$\times$U(1) gauge group.
An important class of such pseudomatter fields, which will be useful here, are the eigenvectors of the following  covariant lattice Lagrangians on a time slice, i.e.
\bea
            -D^{ab}_{\bx \by}[\tU] \xi_n^b(\by;\tU) &=& \l_n \xi_n^a(\bx;\tU) \non \\
            -D_{\bx \by}[V] \r_n(\by;V) &=& \b_n \r_n(\bx;V) \ ,
\label{pseudo}
\eea
 where
\bea
    D^{ab}_{\vx \vy}[\tU] &=& \sum_{k=1}^3 \left[2 \d^{ab} \d_{\vx \vy} - \tU_k^{ab}(\vx) \d_{\vy,\vx+\hat{k}}  - \tU_k^{\dg ab}(\vx-\hat{k}) \d_{\vy,\vx-\hat{k}}   \right]   \non \\
   D_{\vx \vy}[V] &=& \sum_{k=1}^3 \left[2 \d_{\vx \vy} - V^2_k(\vx) \d_{\vy,\vx+\hat{k}}  - V_k^{2\dg}(\vx-\hat{k}) \d_{\vy,\vx-\hat{k}}   \right] \ ,
\label{Laplace}
\eea
and upper Latin indices represent SU(2) colors.  Note that $V$ is squared in $D_{\vx \vy}[V]$.  It is straightforward to verify that the $\rho_n(\bx;V)$ and 
$\xi_n(\bx;\tU$ transform covariantly, like matter fields, under local U(1) and SU(2)$\times$U(1) gauge transformations, 
respectively.  Both are, however, invariant under  transformations in the global center subgroup.  In practice, on the lattice, pseudomatter fields are eigenvectors of large sparse matrices and can be constructed numerically, for given $U,V$ lattice gauge fields, by standard software packages.

\section{\label{states} Physical states}

   In this section we construct locally gauge invariant states for electrically charged and neutral states, of
the following form:
\beq
         \Psi_x = P[x;U,V] M(x;\phi,\psi) \Psi_0 \ .
\eeq
$M(x,\phi,\psi)$ refers to {\it local} composite operators constructed from the fermion and Higgs fields at point $x$, while $P[x;U,V]$ is a pseudomatter field, i.e.\ a non-local functional of the gauge fields alone which transforms locally at the point $x$.  The matter operator $M(x,\phi,\psi)$ may be either gauge covariant or gauge invariant.\footnote{We are concerned here with fermionic particle states.  For gauge bosons, the operator $M$ must include local gauge field operators, see \cite{tHooft:1979yoe,Banks:1979fi,Frohlich:1981yi} and below.   For a general treatment regarding the dressing of local operators to form gauge invariant expressions, cf.\ \cite{Ravera:2026vwp}. }
If gauge invariant, then the state is neutral, and the pseudovector field $P[x;U,V]$ is dispensed with. If 
gauge covariant then the local matter fields source the gauge fields, and the pseudomatter attachment $P(x;U,V)$ is essential for local gauge invariance.  We describe state as electrically charged or electrically neutral according to the transformation properties of the matter operator $M$ under the electromagnetic subgroup, generated by  $T_3 + \oh Y$.

Some of the matter operators for the physical electron and neutrino states in the electroweak theory were discussed long ago \cite{tHooft:1979yoe,Banks:1979fi,Frohlich:1981yi}, 
although  the pseudomatter attachment was never specified, and 
left implicit. In the present work the fermion field $\psi$ transforms in the same SU(2)$\times$U(1) representation as the Higgs field, whereas in the Standard Model the leptons and the Higgs fields have opposite hypercharge, and this leads to some minor differences in the matter operator.   That is not so important, and by reversing the hypercharge of the fermion (which changes the model) the old constructions are recovered.  The  $I_3, Y$ charge assignments used here are shown in Table I.
\begin{table}[htb]
\begin{center}
\begin{tabular}{|c|c|c|c|c|c|c|} \hline
   field component   &   $\psi^1$    &   $\psi^2$   & $\phi^1$ & $\phi^2$ & $\tphi^1$ & $\tphi^2$ \\ \hline
    $I_3$                  &   $\oh$        &   $-\oh$       &  $\oh$    & $-\oh$    & $\oh$ & $-\oh$ \\ \hline   
    $Y$                    &       1            &       1           &  1           &        1     &  -1      &  -1  \\  \hline
\end{tabular}             
\label{T1}
\end{center}
\caption{Hypercharge and the $I_3$ component of isospin in this model.  Note that $\psi,\phi$ in this model have
the same hypercharge.}
\end{table}  

Let us define 
\beq
          \vphi(x) = {\phi(x) \over |\phi(x)|} \ ,
\label{vphi}
\eeq
and also
\beq
         \tvp(x) = i\s_2 \vphi^* \ ,
\eeq 
which transforms like $\vphi$ under SU(2) transformations, but has opposite hypercharge.
The transformation to unitary gauge, belonging to the SU(2) group, is
\beq
    g_{ug}(x;\phi) = \left[ \begin{array}{cc}
                       \vphi^2(x) & -\vphi^1(x) \cr
                       \vphi^{1*}(x) & \vphi^{2*}(x) \end{array} \right] \ ,
\eeq
under which
\beq
    \vphi(x) \rightarrow \left[ \begin{array}{c} 0 \cr 1 \end{array} \right]  ~~~\text{and} ~~~
    \tvp \rightarrow \left[ \begin{array}{c} 1 \cr 0 \end{array} \right] \ ,
\eeq
where the symbol ``$\ra$'', from here through eq.\ \rf{N2}, refers to transformation to unitary gauge.

We begin with the physical state of the neutral fermion \cite{tHooft:1979yoe,Banks:1979fi,Frohlich:1981yi}. Here I will indicate isospin contractions explicitly via upper Latin indices, with the spinor index implicit:
\bea 
             \Psi^{\text{I}}_{\text{neutral}}(x)&=& \vphi^{\dg a}(x) \psi^a(x) \Psi_0\non \\
                         &\ra& \psi^2(x)  \Psi_0 \ ,
\eea
which has electric charge $q=0$, and there is no need for an associated pseudomatter field.
 The upper index ``I''
is introduced to distinguish this state from a different neutral state below.  A physical state of a charged
static fermion is \cite{tHooft:1979yoe,Banks:1979fi}\footnote{Under a U(1) transformation $g_{U(1)}=e^{i\a(x)/2}$, while $\rho(x,V)$ is multiplied by $e^{i\a(x)}$ rather than $e^{i\a(x)/2}$ because it is the square of the link field which appears in the abelian  $D_{xy}$ of \rf{Laplace}.} 
\bea
           \Psi^{\text{I}}_{\text{chrg},n}(x) &=& \r^\dg_n[x;V] \tvp^{\dg a}(x) \psi^a(x) \Psi_0 \non \\
                                                                       &\ra&  \r^\dg_n[x;V] \psi^1(x) \Psi_0 \ .
\eea  
The matter field $ \tvp^{\dg a}(x) \psi^a(x)$, proposed by `t Hooft and by Banks and Rabinovichi  \cite{tHooft:1979yoe,Banks:1979fi}, has electric charge $q=1$, and therefore requires an appropriate 
pseudomatter field.  But these type I states do not exhaust the possibilities.  In fact there is another type of charged state\footnote{States of this type were also employed in \cite{Greensite:2023qfx}, in a related study of SU(3) gauge Higgs theory.}
\bea
       \Psi^{\text{II}}_{\text{chrg,n}}(x) &=&  \xi_n^{\dg b}(\bx,\tU)(\d^{ab} - \vphi^a(x)\vphi^{\dg b}(x))
       \psi(x)\Psi_0 \non \\
        &\ra&  \xi^{1\dg}_n(x,\tU)\psi^1(x) \Psi_0 \ ,
\eea
with charge $q=1$, and also a neutral state
\bea
   \Psi^{\text{II}}_{\text{neutral,n}}(x) &=&  \xi_n^{a\dg}(\bx,\tU)  \vph^a(x) \vph^{\dg b}(x)\psi^b(x)\Psi_0 \non \\
                            &\ra& \xi_n^{\dg 2}(x;\tU) \psi^2(x) \Psi_0 \ ,
\label{N2}
\eea
where both $\pbar^2$ and $\xi_n^2$ are invariant under the electromagnetic gauge group. 
Apart from the type I neutral state, the matter operator $M$ transforms covariantly under $SU(2)\times U(1)$ gauge transformations,
so gauge invariance requires attachment to a pseudomatter field, 
either $\rho_n(x;V)$ for type I charged states, or 
$\xi_n(x,\tU)$ for type II states.  We emphasize again that we assign charge $q$ only to the local composite operator $M(x,\psi(x),\phi(x))$ which is independent of the gauge fields. It is the pseudomatter fields 
$P(x;\tU)$, which depend on the gauge fields at all points in space, that dress the source, restore local gauge invariance, and create a surrounding gauge field.

It is interesting that, allowing for the difference in hypercharge assignments, it was the matter operator for the type II charged fermion state that was chosen in \cite{Frohlich:1981yi} to represent all charged states, as opposed to the operator for type I charged states chosen in \cite{tHooft:1979yoe,Banks:1979fi} for the same purpose.  In fact both choices are valid, but, as we will see, inequivalent.

    As argued elsewhere (\cite{Greensite:2020nhg,Greensite:2021fyi} and in the Appendix), the global center of the gauge group of a gauge Higgs theory is spontaneously broken in the Higgs phase.  However, in the SU(2)$\times$U(1) gauge Higgs theory, the ground state is still invariant under global transformations in the electromagnetic subgroup, which does not contain the center of SU(2)$\times$U(1).  But the operators
 $P[x;U,V] M(x;\phi,\psi)$, while invariant under all local transformations, do transform under the global subgroup of the electromagnetic gauge group, and this has consequences for orthogonality.  The fact that charged states, whether type I or type II, are orthogonal to neutral states is fairly obvious in unitary gauge, since the overlap of such states involves the contraction of the fermion operators $\pbar^a(x) \psi^b(x)$ with $a\ne b$, which vanishes.  But in fact the type I and type II neutral states
are also orthogonal, as are the corresponding charged states.  To see this, consider the product of local SU(2) gauge transformations $g_{SU(2)} = e^{i\a(x)/2}$ with U(1) gauge transformations $g_{U(1)}=e^{i\a(x)/2}$.  For quantities which transform under the fundamental representation of both groups, the combined transformation is
\bea
            G(\a(x)) &=& e^{i\a(x) \s_3} e^{i\a(x)/2} \non \\
                &=& \left[ \begin{array}{cc}  e^{i\a(x)} & 0 \cr 0  & 1 \end{array} \right] \ ,
\label{Galf}
\eea
and under this transformation the color components of $\psi, \xi$ transform in this way: 
\bea
  \psi^1(x) \ra e^{i\a(x)} \psi^1(x)  &~,~& \psi^2(x) \ra \psi^2(x)  \non \\
  \xi^1(x;\tU) \ra  e^{i\a(x)} \xi^1(x;U) &~,~&   \xi^2(x,\tU) \ra  \xi^2(x;U) \ ,
\label{remnant}
\eea
while $\rho(x;V)$ transforms only under $g_{U(1)}=e^{i\a(x)/2}$ as
\beq
\rho(x;V) \ra e^{i\a(x)} \rho(x,V) \ .
\eeq
The scalar $\phi(x)$ transforms like $\psi$, and is unchanged in unitary gauge since  $\phi^1=0$.

Now consider $\a(x)=\a$ independent of space on a time slice.  Under this global subgroup of the electromagnetic subgroup, the transformations of $\phi,\psi$ are the same as above, but, since $V$ is
invariant under global $g_{U(1)}$, so is $\rho(x;V)$ while
\bea
            \xi(x;UV) &\ra&\xi(x;(g_{SU(2)}\circ U)V) \non \\
                           &=& e^{i\a\s_3} \xi(x;UV) \non \\
          \left[ \begin{array}{c}
                \xi^1 \cr \xi^2 \end{array}\right]   &=&   \left[ \begin{array}{c}
                e^{i\a/2} \xi^1 \cr e^{-i\a/2} \xi^2 \end{array}\right] \ ,
 \eea                          
which differs from the second line of \rf{remnant}.
Because the electromagnetic gauge group is unbroken, the ground state is invariant under these global transformations of the fields.  The upshot is that under these global symmetry transformations
\bea
\Psi^{I}(x)_{\text{neutral}} & &\text{is invariant} \non \\
 \Psi^{\text{I}}_{\text{chrg},n}(x) &\ra& e^{i\a}  \Psi^{\text{I}}_{\text{chrg},n}(x) \non \\
\Psi^{\text{II}}_{\text{chrg,n}}(x) &\ra&  e^{i\a/2}  \Psi^{\text{II}}_{\text{chrg,n}}(x) \non \\
\Psi^{\text{II}}_{\text{neutral,n}}(x)  &\ra& e^{i\a/2}  \Psi^{\text{II}}_{\text{neutral,n}}(x) \ .
\label{global}
\eea
From \rf{global} it is clear that the electric charge of the matter source $M(x;\phi,\psi)$ at point $x$ is not determined by the transformation properties of the physical state under U$_\text{EM}$(1).  The differing transformation properties of the last three states under global U$_\text{EM}$(1) can be entirely attributed to the invariance of $V$ under the global transformations, and the effect this has on the global transformation of the pseudomatter fields.   But because of this behavior under a global subgroup of the electromagnetic gauge group, all type I charged states are orthogonal to all type II charged states,  and the type I neutral state is orthogonal to all type II neutral states.  And, as shown above, all neutral states are orthogonal to all charged states.  

    We will see in section \ref{results} that despite their different transformation properties under U$_\text{EM}$(1), neutral type I and type II particles have about the same masses, charged type I and type II particles have about the same spectrum, and charged particles of either type are significantly more massive than neutral particles of either type.

\subsection{Gauge Bosons}

As a side remark, it is worth writing down gauge-invariant operators which create charged and neutral gauge bosons.
For the charged bosons we have
\bea
       \rho_D(x;V)  \vphi^\dg(x) U_\m(x) V_\m^\dg(x) \tvp(x+\hat{\m})   &  &   (W^-) \non \\
       \rho^\dg_D(x;V)  \tvp^\dg(x) U_\m(x) V_\m \vphi(x+\hat{\m})       &  &    (W^+) 
\label{pole}
\eea
Each operator is associated with a Dirac pseudomatter field $\r_D(x;V)$ of the form shown in \rf{rho}, appropriate to the abelian $V$ field,  creating the  corresponding Coulombic field.  Likewise there are two neutral operators   
\bea
        &\vphi^\dg(x) U_\m(x) V_\m(x) \vphi(x+\hat{m}) & \label{nopole} \\
        &\rho_D(x;V) \vphi^\dg(x) U_\m(x) V^\dg_\m(x) \vphi(x+\hat{\m}) \rho^\dg_D(x+\hat{\m})&
        \label{dipole}
\eea
which can be combined to form the $Z$ and photon.  At a glance, the pseudomatter combination $\rho_D(x,V) \rho^\dg_D(x+\hat{\m};V)$ in \rf{dipole} appears to create an electric dipole field.  This is, however, a lattice artifact, since the
electric dipole moment is proportional to the lattice spacing, and disappears in the continuum limit.\footnote{Although this is an old topic, I am not aware that  the expression \rf{dipole} (as opposed to \rf{nopole}) has appeared previously in the literature. Different gauge invariant operators creating both of the neutral gauge bosons have been offered by  Fr{\"o}hlich \cite{Frohlich:1981yi}  and by Maas \cite{Maas:2019nso}, while lattice expressions for the gauge boson operators in unitary gauge are found in Zubkov and Veselov \cite{Zubkov:2008gi}. Presumably in the continuum limit (if it exists) these would all be equivalent.}

\subsection{Time correlators of particle-antiparticle pairs}

   Up to now we have implicitly assumed an infinite volume.  The intuitive                                                                                                                                                                                                                                                                                                                                                                                                                                                                                                                                                                                                                                                                                                                                                                                                                                                                                                                                                                                                                                                                                                                                                                                                                   reason, for charged states, is that it is impossible to insert an electric charge in a finite periodic volume, and although it is possible to formally construct such states, as above, they cannot propagate.  The type I neutral state, which can propagate
freely in a finite periodic volume, is the exception.  These expectations                                                                                                                                                                                                                                                                                       can be verified numerically, as illustrated
in the next section.

    Masses are typically computed from time correlators of a set of gauge invariant operators ${\cal O}$ creating states with the same quantum numbers\footnote{Evaluating the time correlators of covariant operators ${\cal O}$ in a fixed physical gauge, defined by some condition $F[U]=0$,  is equivalent to contracting those operators with pseudomatter fields, since the transformation $g_F(x;U)$ to the gauge is readily checked to have the properties of a pseudomatter field. Then ${\cal O}'(x) = g_F(x;U)\circ{\cal O}(x)$ is a locally gauge invariant operator, and evaluating the time correlator in the $F$-gauge is equivalent to evaluating the time correlator of the ${\cal O'}$ fields with
no constraint in the path integral.}
\beq
           G_{nm}(t) = \langle {\cal O}^\dg_m(t) {\cal O}_n(0) \rangle
\eeq
via the generalized eigenvalue method \cite{Luscher:1990ck,Blossier:2009kd}. On general
grounds
\beq
G_{nn}(t) = \sum_k a_k e^{-E_kt} \ ,
\eeq
where $E_k$ is the energy above the vacuum energy of the k-th energy eigenstate with appropriate quantum numbers. In general $G_{nn}(t)$ converges to $a_0 e^{iE_0(t)}$ at large $t$, where $E_0$ is the minimal energy of a state with these quantum numbers above the vacuum energy.  The exception is if some initial states have very little overlap with the minimal energy state, in which case it may be that $G_{nn}(t)$ converges instead to  $a_j e^{iE_kjt}$ for some $j\ne 0$ for a substantial time interval.  Although this situation may be unusual, it will be relevant below.

    Since electrically charged states can only be created in particle-antiparticle pairs in a finite periodic
volume, we are led to consider a fully gauge invariant state containing static fermion and antifermion particles.
Let us label operators on a time slice $t$
\beq
\Q_A(\bx,t) = \left\{ \begin{array}{cl}
                  \Q             & A  \cr \hline
                  \phi(\bx,t) & N1 \cr
                  \tphi(\bx,t)  \r_n(\bx,t) & C1(n)\cr
                  \vphi(\bx,t) \vphi^\dg(\bx,t))\xi_n(\bx,t) & N2(n) \cr
                  (\mathbb{1} - \vphi(\bx,t)\vphi^\dg(\bx,t))\xi_n(\bx,t) & C2(n)
                  \end{array} \right. \ ,
\eeq
where $N,C$ refer to neutral and charged, respectively, integers 1 and 2 refer to type I and type II states, and lowercase $n$ refers to the pseudomatter fields $\r_n,\xi_n$.  Define the timelike Wilson line
\beq
W_t(\bx,t_0) = U_4(\bx,t_0) U_4(\bx,t_0+1)...U_4(\bx,t_0+t-1) \ .
\eeq
 
We thus consider the time correlation function for a static fermion-antifermion pair (AA$\ra$BB)
\bea
    C_{AB}(t) &=&   \bigg\langle [Q_B^\dg(\bx,t) \psi(\bx,t) \pbar(\bx,0) Q_A(\bx,0)] \non \\
     & &   \times     [Q_A^\dg(\by,0)\psi(\by,0) \pbar(\by,t) Q_B(\by,t)]\bigg\rangle \ .
\eea
By ``well separated'' on a finite lattice we choose $|\bx-\by|=L/2$, where $L$ is the spatial extension,
and adopt the notation that a sum over $\{\bx \by\}$ represents a sum over all sites separated by $L/2$
lattice spacings in the x, y, or z directions.
Integrating out the fermion fields and dropping an irrelevant factor of $(2\k)^{2t}$ (which simply provides an
uninteresting lattice bare mass), the time correlator of interest, averaged over space and 
 $L_4=60$ time slices, is
\bea
C_{AB}(t) &=& {1\over 6 V}\sum_{\{\bx\by\}} \sum_{t_0=0}^{L_4-1}  
\bigg\langle  [Q^\dg_A(\by,t_0) W_t(\by,t_0) Q_B(\by,t_0+t)] ]  \non \\
& & \qquad\ [Q_B^\dg(\bx,t_0+t) W^\dg_t(\bx,t_0)Q_A(\bx,t_0) \bigg\rangle \ ,
\eea
where $V$ is the lattice volume, and $t_0+t$ is taken modulus the $L_4$ time extension to preserve time periodicity. In words, this is just the expectation value of a Wilson line of length $t$
contracted with $Q_A,Q_B^\dg$ operators at the end, times the conjugate operator separated by $\oh L$ lattice spacings in a space direction.   $C_{AB}(t)$ can be evaluated by Monte Carlo simulation, with results described in the next section.  We will also need the single particle time correlation function
\bea
C_{A}(t) &=& {1\over V}\sum_{\bx} \sum_{t_0=0}^{L_4-1}  \bigg\langle Q^\dg_A(\bx,t_0) W_t(\bx,t_0) Q_A(\bx,t_0+t) \bigg\rangle \ . \non \\
\eea

\section{\label{results}Results}

\begin{figure}[t!]
 \includegraphics[scale=0.55]{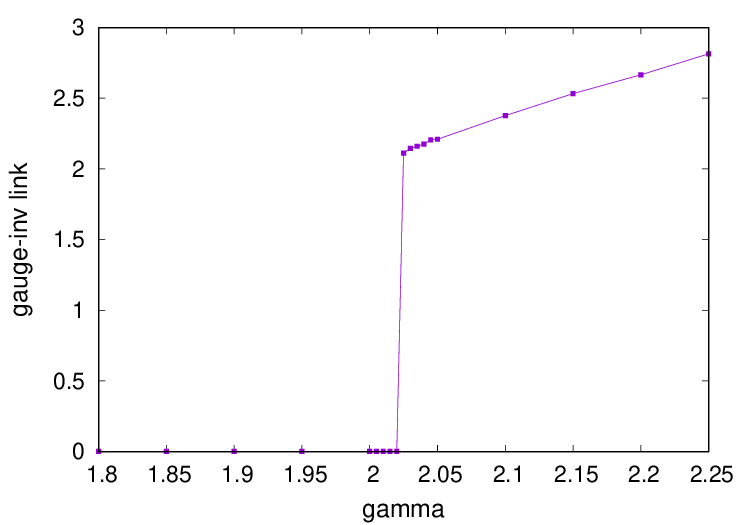}
 \caption{A plot of the expectation value of the gauge invariant link operator $\langle \phi^\dg(x) U_\m(x) \phi(x+\hat{\m})\rangle$ vs.\ $\g$, at fixed  $\b_1=\b_2=3, \lambda=0.13$, averaged over the $12^4$ lattice volume.  There seems to be a very strong first order transition from the symmetric to the Higgs phase near 
 $\g=2.02$.
 }
 \label{phase}
 \end{figure}

     It is important to carry out simulations in the Higgs phase of the SU(2)$\times$U(1) lattice theory, and since for now we are not constrained to simulate the realistic electroweak theory, we can
make an arbitrary choice of couplings, $\b_1=\b_2=3, \g=2.1, \l=0.13$.  It is necessary to check, however, that this point in phase space is in the Higgs phase.  One way to do this, in
the absence of a thermodynamic transition, is to use the non-local order parameter described in the
Appendix \ref{A1} and elsewhere, and in fact this procedure was carried out in \cite{Gangwani:2023dye} for an SU(2)$\times$U(1) gauge Higgs theory with a unimodular Higgs field and parameter choices different from those here.  But a thermodynamic transition separating the symmetric and Higgs phases, where it exists, is easier to spot.   Fig.\ \ref{phase} shows the observable ${\langle \phi^\dg(x) U_\m(x) \phi(x+\hat{\m})\rangle}$, vs.\ the parameter $\g$ on a  $12^4$ lattice with the other couplings above held fixed.   While we have not attempted a careful finite volume analysis, it seems fairly clear (just by eye) that there is a strong first order transition, from the symmetric phase to the Higgs phase, around  $\g=2.02$.  In fact the thermodynamic phase structure of the SU(2)$\times$U(1) gauge-Higgs theory was displayed long ago by Shrock \cite{Shrock:1985ur}, and the fact that this particular choice of couplings is in the Higgs phase is consistent with those results.
 
\begin{figure}[htb]
 \subfigure[~]{
 \includegraphics[scale=0.55]{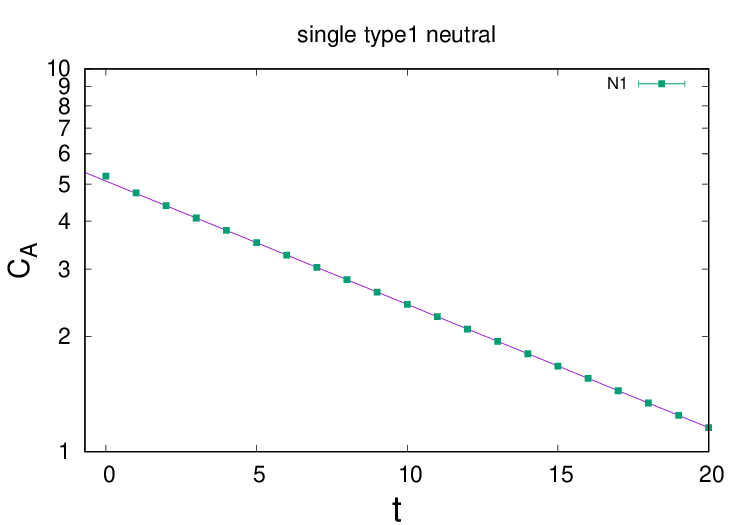}
 \label{singlen1}
 }
\subfigure[~]{
 \includegraphics[scale=0.55]{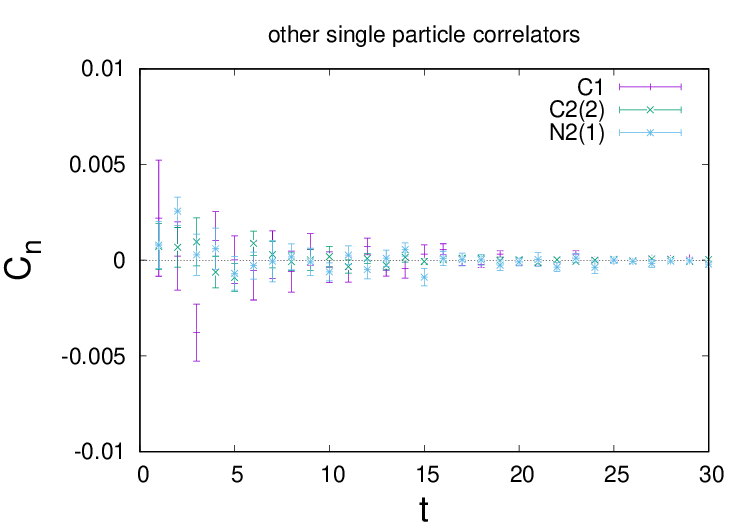}
 \label{singleA}
 }
 \caption{Single particle $A\ra A$ time correlators $C_A$, the keys show the $A$ values of data points.
 (a) Log plot of the time correlator of the neutral type I state.  The energy derived is, within errors, equal
 to half the energy of derived from the corresponding neutral type I particle-antiparticle time correlators shown below.
 (b) No isolated particle apart from type I neutral can propagate; the time correlator of all others must be zero.  This seems to be the case (up to errorbars) in the data shown for charge types I and II, and neutral type II.}
 \label{single}
 \end{figure}

   All other plots were obtained from data obtained on $12^3\times 60$ lattice volumes; energies obtained on other volumes by the same methods are shown in Table I below.\footnote{Correlators were obtained  from lattice Monte Carlo simulations, with data-taking sweeps separated by 100 lattice update sweeps after 10,000 thermalizing sweeps  The total number of data-taking sweeps on
each of the $L^3\times 60$ lattices was 2791,1283, 826, 455 for $L=8,10,12,14$ respectively, corresponding to 16 hours cpu time each on a local cluster.}   Since we work on a finite periodic lattice, it is impossible to create a single charged particle state which propagates in time; this holds true for both type I and type II charged states.  Type II neutral single particle states also cannot propagate, since we have seen that, despite their neutrality, they still transform covariantly under a certain global symmetry \rf{global}.  Fig.\ \ref{single} is just a confirmation of this fact.
The time propagator, denoted $C_A(t)$ for the type I neutral particle is shown on a logarithmic scale in 
Fig. \ref{singlen1}, and from a single exponential fit we obtain a mass in lattice units of 0.07416(1).  In contrast, the time correlators for charged type I, charged type II (n=2), and neutral type II (n=1), are shown on a linear scale in Fig.\ \ref{singleA}, and in these cases $C_{A}$ is consistent with zero, as expected.

The next figure, Fig.\ \ref{C123}, contains the main messages of this article.  Time correlators $C_{AA}(t)$
are displayed for:
\begin{itemize}
\item the type I neutral state ($A=N1$), energy = 0.1483(2).
\item the type I charged stated ($A=C1$), energy = 0.281(4).
\item the type II charged state using the $\xi_2$ pseudomatter field ($A=C2(2)$, energy = 0.278(4).
\item an excited type II charged state using the $\xi_{10}$ pseudomatter field ($A=C2(10)$, energy=0.337(7).
\end{itemize}
In order to aid in a visual comparison of the $C_{AA}(t)$ data on a logarithmic plot, I have rescaled each 
of the $C_{AA}(t)$ data sets so that they all agree at $t=3$.
The quoted energies are obtained from one-exponential fits to the data (also shown) in the range $5\le t\le 15$, where the data has a very nearly linear falloff on a logarithmic plot, and refer to the energies of the particle-antiparticle pair above the bare mass.  The neutral pair is, unsurprisingly, significantly lower in energy than any of the charged pairs. It is also double the energy derived for a single neutral particle.  The type I charged state and some of the type II charged states ($n=2$ is shown) have very nearly the same energies, but this is not the case for larger $n$, such as $n=10$ shown, which 
implies the existence of a charged particle spectrum.  The energy difference between C2(2) and C2(10), which is far outside error bars, cannot be accounted for by  photons in the more energetic state, since even a single photon on a periodic lattice with a spatial extension of 12 lattice spacings would have a minimal non-zero energy of 0.524 in lattice units. This is very far above the energy differences among charged states seen here.

We have suggested elsewhere, for other theories, that ``elementary'' particles in a gauge Higgs theory might have an excitation spectrum; this was previously studied by the author for SU(3) gauge Higgs theory \cite{Greensite:2020lmh}, for the abelian Higgs model by Matsuyama \cite{Matsuyama:2020tvt}, and for the Landau-Ginzburg model of ordinary superconductivity by Matsuyama and myself \cite{Greensite:2022zql} .  The SU(2)$\times$U(1) case studied here is another example.  We also note some very recent work by Martins et al.\ \cite{Martins:2026eni} which proposes to look for particle spectra of this type by other means.

\begin{figure}[t!]
 \includegraphics[scale=0.65]{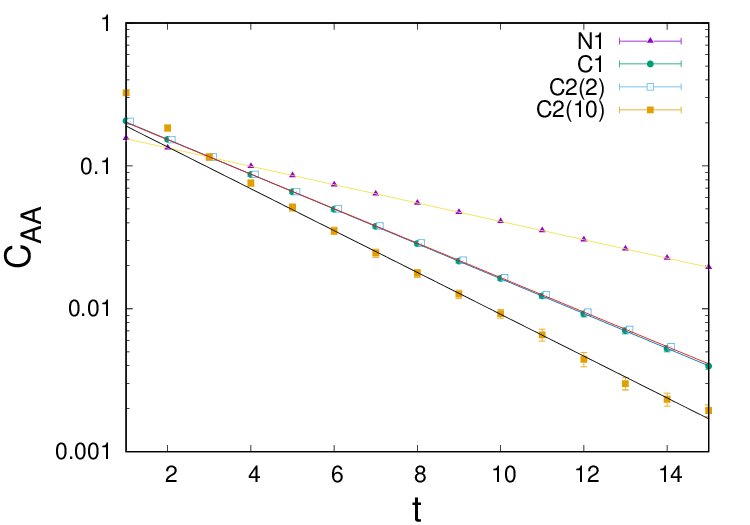}
 \caption{Logarithmic plot of the $AA\ra AA$ time correlator for a static fermion-antifermion pair, for a  type I neutral pair, a type 1 charged pair, and two type II charged pairs with pseudomatter $\xi_2$  ($n=2$) and 
$\xi_{10}$ ($n=10$) respectively.  The data was obtained on a $12^3\times 60$ lattice volume, and the static particles are separated by six lattice spacings.}
 \label{C123}
\end{figure}

It was noted earlier that time correlation functions such as $C_{AA}(t)$ generally converge to the exponential falloff of the lowest state with the quantum numbers of the initial and final states.  The exception is for states
whose overlap with the ground state is very small, and therefore falls off exponentially with the energy of
a single excited state, at least for some large range of $t$.  This appears to be the case for the $C_{AA}(t)$ correlator with $A=C2(10)$ shown in Fig.\ \ref{C123}.   In the absence of such exceptional situations, the generalized eigenvalue approach \cite{Luscher:1990ck,Blossier:2009kd} is used.  In brief, the idea is as follows:  In order to compute the $N$ lowest energy eigenvalues in the spectrum, define a matrix $C(t)$ whose components are $C_{AB}(t)$ with indices $A,B$ running over $N$ chosen operators.  The generalized eigenvalue equation is
\beq
           C(t) \vec{v}_n(t,t_0) = \l_n(t,t_0) C(t_0) \vec{v}_n(t,t_0) \ .
\eeq
Then, with various caveats about sources of error (see \cite{Luscher:1990ck,Blossier:2009kd})
\bea
E_n &=& \lim_{t\ra \infty} E_n^{eff}(t,t_0) \non \\
E_n^{eff}(t,t_0) &=& -\log\left[{\l_n(t+1,t_0)\over \l_n(t,t_0)}\right] \ .
\eea

\begin{figure}[t!]
 \includegraphics[scale=0.65]{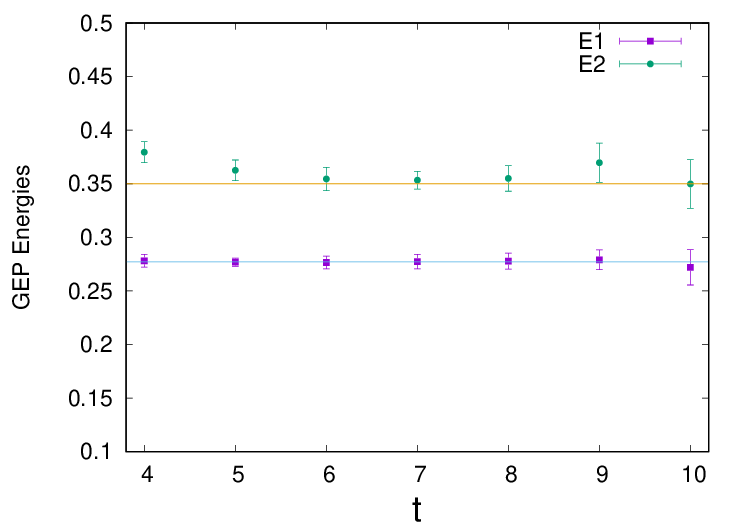}
 \caption{The two lowest energies of the type II charged pairs, via the generalized eigenvalue method.}
 \label{gepeig}
\end{figure}

      In order to carry out this procedure over a substantial range of $t-t_0$, very accurate data is required.
So in order to find a ground state and first excited state of the charged type II states, we set $N=2$ and use the $C_{AB}(t)$ correlators with the most accurate data, which correspond to $A,B = C2(2), C2(3)$. 
The results for energies $E_n^{eff}(t,t_0)$ are shown in Fig.\ \ref{gepeig}, at $t_0=3$.  Error bars are estimated by applying the generalized eigenvalue procedure to the correlators obtained from each of ten independent runs, and the energies are consistent (or perhaps slightly higher for the first excitation) with those obtained from the simple exponential fits to the data shown in Fig.\ \ref{C123}.  The results are fairly insensitive to the choice of $t_0$.  At higher
$N$ the lowest two energies are still roughly consistent with $N=2$, but error bars become very large. 

\begin{widetext}
\begin{table*}[htb]
\begin{center}
\begin{tabular}{|c|c|c|c|c|c|c|c|} \hline
  lattice/state                       &       C1          &     C2(2)    &     C2(4)      &     C2(6)       &   C2(8)      &    C2(10)   &  C2(12)  \\ \hline
   $8^3\times 60$        &  0. 274(2)   &  0.275(2)     &  0.331(6)   & 0.336(9)   &  0.344(11)   &  0.338(11)  & 0.347(5)   \\ \hline 
   $10^3\times 60$      &  0. 277(3)   &  0.278(4)     &  0.321(21)   & 0.315(12)   &  0.328(7)   &  0.338(14)  & 0.340(10) \\  \hline 
   $12^3\times 60$      &  0. 281(4)   &  0.278(4)   &  0.309(20)   & 0.300(12)   &  0.307(10)   &  0.337(7)  & 0.332(20)   \\ \hline
   $14^3\times 60$      &  0. 277(3)   &  0.277(4 )    &  0.297(35)   & 0.303(25) &  0.305(21) &  0.323(7)  & 0.345(9)   \\ \hline  
\end{tabular}
\label{T2}
\end{center}
\caption{Energies of charged static fermion-antifermion pairs above the bare mass value, separated by half the space extension of the lattice, on lattices of differing spatial volumes.}
\end{table*} 
 \end{widetext}
  
   The energies and errors for type II charged states extracted from one-exponential fits to time correlators in the time range $5\le t\le 15$, at various lattice volumes, are listed for a selection of C2(n) states in Table II. To avoid confusion, it should be understood that there is no question of using this data for scattering, since we are looking at two static particles at fixed positions.  Neither are we looking for bound state energies; the idea of separating the particles by half the spatial lattice extension is to try to minimize the effect of interactions.  The only reason for considering time correlators of particle-antiparticle pairs is because, for single particle states which transform covariantly under the remnant global symmetry \rf{remnant}, the time correlator is zero, as seen in Fig.\ \ref{single}.   Therefore we need a pair of particles to restore invariance; the particle-antiparticle separation is unimportant for that purpose.  Taking account of error bars, the data indicates a lowest energy of the particle pair at around 0.277 in lattice units, and at least  one excited state energy around 0.33-0.34 in lattice units.  It is possible that there are still higher excitations in the $C2(n)$ spectrum at higher $n$.  Numerically, however, the error bars in the 
$C_{AA}(t)$ data grow with $n>14$ or so, and a fit in the $5\le t\le 15$ range becomes unreliable.

\begin{figure}[htb]
 \subfigure[~]{
 \includegraphics[scale=0.65]{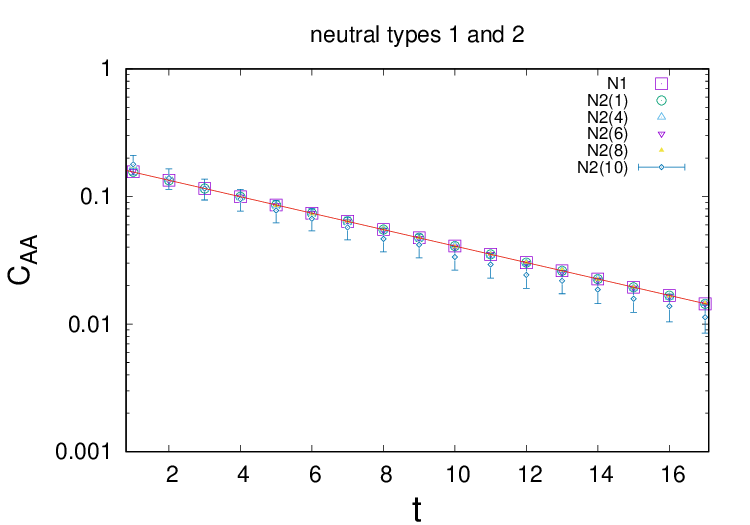}
 \label{neutral}
}
\subfigure[~]{
\includegraphics[scale=0.65]{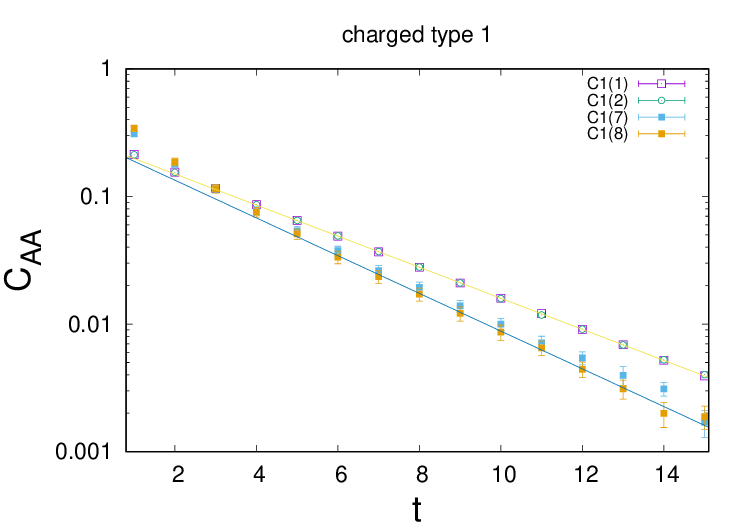}
\label{chrg1}
}
\caption{A selection of time correlators for (a) neutral type I (N1) and neutral type II fermion pairs, and
(b) charged type I fermions, using pseudovectors $\r_n(x,V)$ defined in \rf{pseudo}.  There is a near perfect overlap between the $n=1$ and $n=2$ data points.}
\end{figure}

   We next consider the energies that can be extracted from the neutral type I and type II correlators.
The results for a sample of such correlators are shown in Fig.\ \ref{neutral}, again rescaled to all agree
at time $t=3$.  The energy derived from the fit to the type I neutral state is 0.1483(2), and this energy is shared by all the type II neutral states shown.  At larger $n$ the error bars become substantial, there are deviations from a near-perfect overlap of data points, but these are mainly within error bars, and there is no convincing indication from the data of excited neutral states.  If there are such excitations, one might consider generalizations of type I neutral states of this kind:
\beq
   \Psi(x) = \sum_y \phi^\dg (y) V(y,x;U;V) \psi(x) \Psi_0 \ .
\eeq
Where $V(y,x,U;V)$ is) some functional of the gauge fields which transforms bi-covariantly at points $y,x$, i.e.\ like a Wilson line (but not necessarily itself a Wilson line) between those points.  In the absence of an obvious choice for $V(y,x;U;V)$ I have not tried to do that here.

   The situation is a little different for charged type I pairs, where we consider the matter field coupled to
pseudovector $\r_n(x;V)$ defined in \rf{pseudo}.  A selection of C1(n) time correlators are shown in Fig.\ \ref{chrg1}, and the straight line fits
are to C1(1) (also shown in Fig.\ \ref{C123}) and to C1(8), which provide mass estimates of  0.280(2) and 0.340(14) respectively.  These are compatible with our previous estimates for the charged type II states, and here again we see an indication of excitations in the spectrum of charged static particles.  The result suggests that type I and type II charged states, which transform differently with respect to the global subgroup of the
electromagnetic group, may have a degenerate spectrum.

\section{\label{conclude}Conclusions}

    It was shown a long time ago that one may construct without gauge fixing, in an SU(2)$\times$U(1) gauge Higgs theory in the Higgs phase, particle creation operators which are either invariant, or which transform covariantly, under the unbroken electromagnetic symmetry \cite{tHooft:1979yoe,Banks:1979fi,Frohlich:1981yi}.\footnote{The one-to-one correspondence between operators in the electroweak Lagrangian and physical particles is a special feature, among non-abelian theories, of the electroweak gauge group, as emphasized by Maas et al.\  \cite{Maas:2017xzh}.}   Left implicit in the early work was that in order to obtain locally gauge invariant states for electrically charged particles,  the corresponding particle creation operators must be attached to (or ``dressed by'') pseudomatter operators.  These are functionals of the gauge fields which, apart from invariance under the global center subgroup of the gauge group, transform like matter fields in a fundamental representation.  It has been shown in this article that there are qualitatively different ways to
introduce both neutral and charged particle operators, denoted type I and type II, and that operators of each
type transform differently under a global subgroup of the electromagnetic gauge group, even if both sets correspond to the same charge.  In this sense there are a variety of electrically charged physical states in SU(2)$\times$U(1) gauge Higgs theory, two kinds of neutral particle operators and two kinds of charged particle operators, invariant under all local gauge transformations, but differing by their transformation properties under a global subgroup of the electromagnetic gauge group.

     To summarize the numerical results:  neutral type I and neutral type II particles have about the  same masses, and these are significantly smaller than the masses of charged type I and charged type II particles.  Both type I and type II  charged particles have a spectrum consisting of at least two masses (perhaps more), and each of these is about the same for the two types.  The two types of charged particles and the two types of neutral particles are orthogonal, and distinguished by their transformation properties under an unbroken global subgroup.
     
     Whether these results have any application to quarks and leptons in the electroweak sector of the Standard Model is an open question. One may conjecture that the fermion excitations seen for static sources, or even the existence of two types of charged/neutral states, might have something to do with particle generations.  But at the moment this is pure speculation. If a practical method for simulating chiral fermions ever becomes available, then it would be interesting to return to this question.

\appendix

\section{\label{A1} Order parameter, and the spin glass analogy}
            
Here we elaborate on comments made at the beginning of section \ref{charge}, to the effect that the expectation value of operators defined at fixed time are equivalent to the expectation of operators in a
certain type of spin glass.

The orthogonality of physical (e.g.\ charged and uncharged) states is determined by their overlap at a fixed time, and this in turn is expressed as the vacuum expectation value of products of operators at equal times.  We therefore consider operators $\Ac$ in a gauge Higgs theory which are functionals of the fields $\bU, \varphi$ on a time slice, invariant under local gauge transformations, but which may transform covariantly under the global center subgroup of the gauge group.  Here we show that $\langle \Ac \rangle$ can be expressed as the expectation of that operator in a certain kind of spin glass.  In a spin glass there is a global symmetry acting on the spins which is realized in either a symmetric or spontaneously broken phase, the latter being the spin glass phase.  In a gauge Higgs theory, we identify the spin glass phase as the Higgs phase. 

Let $\bU_k(\bx) = U_k(\bx,0), k=1-3$ and $\varphi(\bx)=\phi(\bx,0)$ represent lattice gauge fields on a time slice $t=0$.\footnote{The symbol $\varphi(\bx)$ used in this appendix should not be confused with the rescaled $\phi(x)$ field defined in eq.\ \rf{vphi} of the main text.}  Define

\bea
\int D\U &\equiv& \int DU_{k>0}(\bx,t\ne 0) DU_0(\bx,t) \non \\
\int DU D\phi &=& \int D\bU D\U D\phi \ .
\eea
Then
\bea
\langle \Ac \rangle &=& {1\over Z} \int DU D\phi \Ac[\bU,\varphi] e^{-S} \non \\
&=& \int D\bU {Z[\bU] \over Z} {1\over Z[\bU]}  \int D\U D\phi \Ac[\bU,\varphi]e^{-S} \ ,
\eea
where
\beq
Z[\bU] = \int D\U D\phi e^{-S} \ .
\eeq
Define
\beq 
P[\bU] = {Z[\bU] \over Z} = {1\over Z}  \int D\U  D\phi e^{-S} \ .
\label{PU}
\eeq
This is a probability distribution for a configuration $\bU$, averaging over all other fields on the lattice weighted by $e^{-S}/Z$, and the identity
$\int D\bU P[\bU]=1$ is immediate. But then
\bea
\langle \Ac \rangle &=& \int D\bU P[\bU] \left\{ {1\over Z[\bU] } \int D\U D\phi 
\Ac[\bU,\varphi] e^{-S} \right\}  \non \\
&=& \int D\bU P(\bU) \langle \Ac \rangle_{\bU} \ ,
\label{spin_glass}
\eea
where 
\beq
 \langle \Ac \rangle_{\bU} = {1\over Z[\bU] } \int D\U D\phi  \Ac[\bU,\varphi] e^{-S} \ .
\eeq
This describes a type of spin glass in which the Higgs field $\varphi$ plays the role of a spin variable,
and link variables $\bU$ are the random couplings with probability distribution $P[\bU]$.  This differs from an ordinary spin glass only in the fact that the link probability distribution $P[\bU]$ is fixed to \rf{PU}; it is non-local and cannot be altered without changing the theory.  The variables at $t\ne 0$ are essentially auxiliary variables which may (in principle) be integrated out,  leaving an expression depending entirely on field values at $t=0$.  Then it is easy to see that, as in a spin glass, the theory with fixed $\bU$ defined by the partition function $Z[\bU]$ is invariant under the transformations $\phi \ra z \phi$, where $z$ is an element of the global center gauge group.  But this global transformation could, in principle, break spontaneously.\footnote{In order to discuss spontaneous symmetry breaking more rigorously, one
must add a small global symmetry breaking term to the action proportional to some parameter $h$, and then carry out the computation of the order parameter in the limits $V \ra \infty$ and $h\ra 0$ taken in that order.  The formulation can
be found in detail in ref.\ \cite{Greensite:2020nhg}; for simplicity that modification is left implicit here.  It turns out to be anyway unnecessary 
in numerical simulations.}

Define
\bea
           \op(\bx;\bU) &=& {1\over Z[\bU]}  \int D\U D\phi \varphi(\bx) e^{-S} \non \\
           \Omega[\bU] &=& {1\over V_s} \sum_\bx |\op(\bx;\bU)| \non \\
            \langle \Om \rangle &=& \int D\bU P[\bU] \Omega[\bU] \ ,
\eea
where $V_s$ is the lattice 3-volume of a time slice.  The order parameter $\langle \Om \rangle$ is modeled after the Edwards-Anderson order parameter for
spin glasses  \cite{Edward_Anderson}, 
and it tests whether the global center subgroup of the gauge symmetry is spontaneously broken, in a case
where $\sum_\bx \op(\bx;\bU)/V_s$ vanishes in general, whether the symmetry is broken or not.
For $\langle \Om \rangle = 0$ the theory is in the unbroken phase, and it implies that $\op(\bx;\bU)=0$ at {\it each} $\bx$,
for any configuration (apart from a set of negligible probability) drawn from the distribution $P[\bU]$.  Conversely, $\langle \Om \rangle > 0$ implies a broken center symmetry, and the system is in the Higgs phase \cite{Greensite:2020nhg}.\footnote{If a gauge Higgs theory has a custodial symmetry which transforms only the scalar field, then $\langle \Om \rangle$ is also an order parameter for the spontaneous breaking of that symmetry.   Thus custodial and global center gauge symmetries break simultaneously in the Higgs phase. When custodial symmetry is absent or explicitly broken, as in the realistic electroweak theory, it is the global center of the subgroup alone which is spontaneously broken in the Higgs phase.}  It is clear from its definition that the order parameter is non-local, since $\op(\bx;\bU)$ is a non-local operator.

In order to calculate $\langle \Om \rangle$ numerically, one may run an ordinary lattice Monte Carlo simulation and, at the start of each data taking sweep, the configuration is that obtained from the previous update sweep.
This means that $\bU$ at $t=0$, which is unaltered throughout the data taking sweep, is effectively
drawn from the probability distribution $P(\bU)$.  The data taking sweep is also a Monte Carlo simulation, but with $\bU$ fixed, and all other variables generated by the usual update procedure are drawn from the
Boltzmann distribution $e^{-S}/Z[\bU]$.  Note that this probability density with $\bU$ fixed
\beq
          \int D\U D\phi {e^{S}  \over Z[\bU]} = 1
\eeq 
sums to unity as expected.  The Monte Carlo sweeps within each data-taking sweep are used to compute
$\Om[\bU]$ at the given $\bU$, which itself arises with probability density $P(\bU)$.  Thus, as in a spin glass, the expectation of site variables on a time slice are computed with fixed link variables on that time slice, which are drawn from a certain probability distribution.  The only important difference is that in a spin glass, aside from a few restrictions, the link variable probability distribution $P(\bU)$ can be chosen almost at will, while in gauge Higgs theories the distribution is fixed.

Examples of numerical calculation of the order parameter in gauge Higgs theories are found in
\cite{Greensite:2020nhg,Gangwani:2023dye,Greensite:2023qfx,Ward:2021qqh,Alles:2025kvd,Alles:2024qha}, 
along with simulation details.

     In the symmetric phase, fermions whose color is entirely screened by the scalar field (such as 
 $\phi^\dg (x)\psi(x)\Psi_0$), and fermions which are sources of all gauge fields in the theory (such as 
 $\xi^\dg (x;U) \psi(x) \Psi_0$) form two orthogonal sets of physical states, with orthogonality enforced by their neutrality or covariance under the global center subgroup. This distinction is lost in the transition to the Higgs phase.  When the symmetric phase is in the ``separation of charge''  (S$_\text{c}$) confined phase (see \cite{Greensite:2017ajx}), then the transition to the Higgs phase is accompanied by physical effects, having to do with a loss of  S$_\text{c}$ confinement in the Higgs phase, and a corresponding loss of linear Regge trajectories in that phase.  For a discussion of these points, cf.\ \cite{*Greensite:2020nhg,Greensite:2021fyi}.  Yet physical states which are either electrically charged or electrically neutral certainly exist in the Higgs phase of SU(2)$\times$U(1) gauge Higgs theory, and clearly these cannot be distinguished by their transformation properties under the global center.  Note, however, that the global subgroup of the electromagnetic gauge group does not actually contain the center subgroup of the full gauge group; thus global U$_{\text{EM}}$(1) can be a symmetry of the theory even if the global center subgroup is broken.   That raises the question of how to construct electrically charged or electrically neutral physical states in the Higgs phase, and of course this is what we have addressed, following up on the pioneering work of \cite{tHooft:1979yoe,Banks:1979fi,Frohlich:1981yi},  in the main text.


     
     It should finally be mentioned that there are other views concerning the transition to the Higgs phase in
gauge Higgs theories with the Higgs in the fundamental representation, which differ from what is presented here, and in the references \cite{Greensite:2020nhg,Greensite:2021fyi}  just cited.  In the condensed matter community, some investigators focus on a transition to the Higgs phase happening only at the boundary of a finite volume \cite{Verresen:2022mcr,Thorngren:2023ple,Chung:2024hsq}, while physics in the bulk is considered continuous (which is true as concerns local thermodynamic observables). Of course the role of the boundary is of great importance in certain condensed matter systems.  In particle physics, however, our attention generally focuses on physical effects in the bulk.  

   In the bulk, it is well known that the Higgs and confined phases of an SU(N) gauge Higgs theory, with the Higgs in the defining representation, are not
entirely isolated from one another by a thermodynamic transition and in fact there is always a path in the phase diagram from one phase to the other which avoids a thermodynamic transition \cite{Osterwalder:1977pc, Fradkin:1978dv}. Hence one might question the use of the word ``phase,'' in this context, at least in the bulk, and indeed some prefer the term ``confined-Higgs'' phase.  But here it is important to note that not all physically relevant phase transitions are associated with a thermodynamic transition, a good example being the Kertesz line between percolating and non-percolating phases in the Ising model with an external field \cite{Kertesz}.  Likewise, the transition between symmetric and Higgs phases in gauge Higgs theories may or may not be accompanied by a thermodynamic transition, but certain non-local properties in the two phases, such as the presence or absence of S$_\text{c}$ confinement and linear Regge trajectories, can change across the transition, with the transition line located by 
the order parameter described here.  And while non-analytic behavior of local operators should only appear, in the thermodynamic limit, across a thermodynamic transition, this statement does not necessarily hold for non-local operators, in particular the order parameter $\Om$ defined above.

\bibliography{sym3}
\end{document}